# On the use of human mobility proxies for modeling epidemics

Michele Tizzoni[1], Paolo Bajardi[2], Adeline Decuyper[3], Guillaume Kon Kam King[4], Christian M. Schneider[5], Vincent Blondel[3], Zbigniew Smoreda[6], Marta C. González[5,7], Vittoria Colizza[8,9,10,*]

[1] Computational Epidemiology Laboratory, Institute for Scientific Interchange (ISI), Torino, Italy
[2] Department of Veterinary Science, University of Turin, Torino, Italy.
[3] ICTEAM Institute, Université Catholique de Louvain, Belgium
[4] CNRS, UMR5558, F-69622 Villeurbanne, France
[5] Department of Civil and Environmental Engineering, Massachusetts Institute of Technology, 77 Massachusetts Avenue, Cambridge, MA 02139, USA
[6] Sociology and Economics of Networks and Services Department, Orange Labs, France
[7] Engineering Systems Division, Massachusetts Institute of Technology, 77 Massachusetts Avenue, Cambridge, MA 02139, USA
[8] INSERM, U707, Paris, France
[9] UPMC Université Paris 06, Faculté de Médecine Pierre et Marie Curie, UMR S 707, Paris, France
[10] Institute for Scientific Interchange (ISI), Torino, Italy

* corresponding author : vittoria.colizza@inserm.fr

## Abstract

**Human mobility is a key component of large-scale spatial-transmission models of infectious diseases. Correctly modeling and quantifying human mobility is critical for improving epidemic control, but may be hindered by data incompleteness or unavailability. Here we explore the opportunity of using proxies for individual mobility to describe commuting flows and predict the diffusion of an influenza-like-illness epidemic. We consider three European countries and the corresponding commuting networks at different resolution scales, obtained from (i) official census surveys, (ii) proxy mobility data extracted from mobile phone call records, and (iii) the radiation model calibrated with census data. Metapopulation models defined on these countries and integrating the different mobility layers are compared in terms of epidemic observables. We show that commuting networks from mobile phone data well capture the empirical commuting patterns, accounting for more than 87% of the total fluxes. The distributions of commuting fluxes per link from mobile phones and census sources are similar and highly correlated, however a systematic overestimation of commuting traffic in the mobile phone data is observed. This leads to epidemics that spread faster than on census commuting networks, once the mobile phone commuting network is considered in the epidemic model, however preserving to a high degree the order of infection of newly affected locations. Proxies' calibration affects the arrival times' agreement across different models, and the observed topological and traffic discrepancies among mobility sources alter the resulting epidemic invasion patterns. Results also suggest that proxies perform differently in approximating commuting patterns for disease spread at different resolution scales, with the radiation model showing higher accuracy than mobile phone data when the seed is central in the network, the**

**opposite being observed for peripheral locations. Proxies should therefore be chosen in light of the desired accuracy for the epidemic situation under study.**

## Author summary

The spatial dissemination of a directly transmitted infectious disease in a population is driven by population movements from one region to another allowing mixing and importation. Public health policy and planning may thus be more accurate if reliable descriptions of population movements can be considered in the epidemic evaluations. Next to census data, generally available in developed countries, alternative solutions can be found to describe population movements where official data is missing. These include mobility models, such as the radiation model, and the analysis of mobile phone activity records providing individual geo-temporal information. Here we explore to what extent mobility proxies, such as mobile phone data or mobility models, can effectively be used in epidemic models for influenza-like-illnesses and how they compare to official census data. By focusing on three European countries, we find that phone data well matches the commuting patterns reported by census but tends to overestimate the number of commuters, leading to a faster diffusion of simulated epidemics. The order of infection of newly infected locations is however well preserved, whereas the pattern of epidemic invasion is captured with higher accuracy by the radiation model for centrally seeded epidemics and by phone proxy for peripherally seeded epidemics.

# Introduction

One of the biggest challenges that modelers have to face when aiming to understand and reproduce the spatial spread of an infectious disease epidemic is to accurately capture population movements between different locations or regions. In developed countries this task is generally facilitated by the existence of data or statistics at the national or regional level tracking individuals' movements and travels, by purpose, mode, and other indicators if available (see e.g. transport statistics in Europe [1], commuting, migration data or other types of mobility at country level [2-6]). Access to highly detailed and updated data may however still be hindered by national privacy regulations, commercial limitations, or publication delays. The situation becomes increasingly complicated in less-developed regions of the world, where routine data collection may not be envisioned at similar levels of details [7], but which, most importantly, may be characterized by a high risk of emergence and importation of infectious disease epidemics or may suffer of endemic diseases.

Depending on the infectious disease under study, different mobility processes may play a relevant role in the spatial propagation of the epidemic while others appear to be negligible, as determined by the typical timescales and mode of transmission of the disease, and the geographic scale of interest. For rapid directly transmitted infections, daily movements of individuals represent the main mean of spatial transmission. At the worldwide scale, air travel appears to be the most relevant factor for dissemination, as observed during the SARS epidemic [8, 9] and the 2009 H1N1 pandemic [10,11]. On smaller regional scales, instead, daily commuting is significantly linked to the spread of seasonal influenza [12,13], affecting the epidemic behavior at the periphery of the airline transportation infrastructure [14].

To overcome issues in accessing commuting data when simulating spatial influenza spread, epidemic models have traditionally relied on mobility models to synthetically build patterns of movements at the desired scale [14-16]. The gravity model [17] and the recently proposed radiation model [18] have been shown to well fit commuting patterns observed in reality on different spatial scales [12,14-16,18-20].

Next to mobility modeling approaches, alternative tools for understanding daily human movements have more recently flourished thanks to the availability of individual data obtained from different sources, namely mobile phone call records carrying temporal and spatial information on the position of the cell phone user at the level of tower signal cells [21-23]. Such direction of research has gained great popularity, leading to the discovery of universal characteristics of individual mobility patterns, and the possibility to study mobility in space and at timescales that were unreachable before [21-26]. Such increasing volumes of finely resolved human mobility data, thanks to the near ubiquity of mobile phones, also offered an opportunity to contrast the huge deficit of quantitative data on individual mobility from underdeveloped regions. They were indeed used to shed light on malaria diffusion and identify hotspot areas [24,26,27], to monitor human displacements in case of natural disasters [25,28] and to study disease containment strategies in Ivory Coast [29].

Despite the variety of modeling approaches and data sources, the impact of using different proxies for human commuting in epidemic models for rapidly disseminated infections is still poorly understood. Each approach or source of data clearly has its own intrinsic strengths and weaknesses, related to accuracy and availability of the dataset.

More specifically, mobility models require some assumptions or input data for calibration and fit to the real commuting behavior. The gravity model requires full knowledge of mobility data for its parameter fitting and can be extended to other regions where data is not available in case of empirical evidence pointing to "universal" commuting behavior at a given resolution scale, i.e. well described by the same set of parameter values [14], or by making assumptions on generalizability. The radiation model requires population distribution values and the total commuter flows out of a given region, a quantity that may not be easily accessible at the desired level of resolution or with sufficient coverage. While mobile phone data can provide mobility information at a high granularity level, they are also characterized by a number of issues that may hinder their use. Phone data are inevitably affected by biases related to the population sampling: coverage is usually not homogenous across space and it depends on the market share of the operator providing the data. Phone ownership and usage may differ across social groups, gender or age classes depending on the country under study [30,31], and access to users' metadata to evaluate the representativeness of the sample is limited by privacy concerns [32]. Given the recent availability of these data, the impact of such biases on mobility estimates is still poorly understood.

Recent studies have assessed the effects of using gravity models in mathematical epidemic models [12,33], however similar works on the use of data-saving options like the radiation models or of alternative strategies like mobile phone activity data for epidemic applications are still missing.

The aim of this paper is therefore to assess the adequacy of two specific proxies – mobile phone data and the radiation model – to reproduce commuter movement data for the modeling of the spatial spread of influenza-like-illness (ILI) epidemics in a set of European countries. We first compare the commuting networks extracted from the official census surveys of three European countries (Portugal, Spain and France) to the corresponding proxy networks extracted from three high-resolution datasets tracking the daily movements of millions of mobile phone users in each country. More specifically, we examine through a detailed statistical analysis the ability of mobile phone data to match the empirical commuting patterns reported by census surveys at different geographic scales. We then examine whether the observed discrepancies between the datasets affect the results of epidemic simulations. To this aim, we compare the outcomes of stochastic SIR epidemics simulated on a metapopulation model for recurrent mobility that is based either on the mobile phone commuting networks or the radiation model commuting networks, with respect to the epidemics simulated by integrating the census data. We evaluate how the simulated epidemic behavior depends on the underlying mobility source and on the spatial resolution scale considered, by investigating the time to first infection in each location and the invasion epidemic paths from the seed.

## Materials and Methods

### Ethics statement

The study relied on billing datasets that were previously recorded by a mobile provider as required by law and billing purposes, and not for the purposes of this project. To safeguard personal privacy, individual phone numbers were anonymized by the operator before leaving storage facilities, in agreement to national regulations on data treatment and privacy issues, and they were identified with a security ID (hash code). The research was reviewed and approved by the MIT's Institutional Review Board (IRB). As part of the IRB review, authors, who handled the data, and the PI participated in ethics training sessions at the outset of the study.

### Commuting networks

**Commuting networks extracted from census surveys**

The census commuting networks are extracted from three census surveys, one for each of the countries under study: Portugal, Spain and France. Each survey tracks the number of people who daily commute for work or study reasons between any two locations within the country. Locations are identified as political subdivisions of the country, usually corresponding to their lowest administrative level. Commuting flows directed to or coming from abroad are not considered in the analysis (see Section 3.1 in Text S1 for a sensitivity analysis on cross-border commuting). Networks are generated by creating a directed weighted link between two nodes, representing the locations of origin and destination, and the weight indicates the number of commuters traveling on that connection on a typical working day.

Census surveys used in the present study are not homogeneous in terms of collection date and geographic resolution at which data is collected. They represent however the most accurate and reliable description of commuter movements that is available for the countries under study. Commuting data for Portugal is extracted from the database of the National Institute of Statistics [5] and refers to the 2001 National Census Survey. The survey is nationwide and data is collected at the level of *freguesia* (namely, parish) that is the smallest local administrative unit of Portugal. For the purpose of making a comparison between census and mobile phones data feasible, we need to coarse-grain the commuting data on larger spatial scales corresponding to higher administrative levels of the country. This ensures the establishment of a resolution level that is common to both approaches and that allows the coarse-graining of each dataset maintaining the quality of the data. Parishes are indeed very heterogeneous in terms of surface area (from 100m$^2$ to 100km$^2$) and are not structured in a hierarchical form with respect to tower cells as the latter can be smaller than a parish or emcompass a full parish.

We thus consider: (i) the Portuguese *concelhos* (roughly corresponding to municipalities and typically including tens of *freguesias*) and (ii) the *distritos* (districts), the largest administrative unit of Portugal. We exclude from our analysis all municipalities located on the islands.

Commuting data for Spain is extracted from the database of the National Institute of Statistics [6] and refers to the national workforce survey for the year 2005. The survey is conducted over a population sample and data is provided at the geographical level of provinces. We project the data to the full population of the country and restrict our analysis to continental Spain only.

Commuting data for France is extracted from the database of the French National Institute of Statistics and Economic Studies [4] and refers to the 2007 National Census Survey. The survey is nationwide and data is collected at the level of *communes,* the smallest local administrative unit of France. Similarly to the case of Portugal, we coarse-grain the original network on two higher administrative levels, corresponding to: (i) the French *arrondissements* (districts) and (ii) the French *departments*. For consistency, we exclude from our analysis all the overseas regions and territories of France.

In the following we indicate with $w_{ij}^c$ the census flux of commuters from the administrative unit $i$ to the administrative unit $j$. In Section 1 of Text S1 we report additional details about the sources and the definitions of the census data.

**Commuting networks extracted from mobile phones records**

Mobile phone commuting networks are extracted from three high-resolution datasets, based on mobile phone's billing information of a large sample of anonymized users in each country under study (2006 data for Portugal, 2007 for Spain and France), and already used in previous works [34-37].

The data provides information about the time of usage of the mobile phone and the coordinates of the corresponding mobile phone tower handling the communication. The data allows us to identify the set of locations visited by each user (georeferenced in terms of tower cells) and to rank them according to the total number of calls placed by a user from each of them. Only users with more than 100 calls are included in the study, to enable the estimation of the individual's commuting mobility pattern. Since mobile phone trajectories clearly include different sorts of daily movements, we need therefore to extract commuter movements only for the comparison with census data, and disregard other types of displacement. Following previous work [22], we assume that a user's residence corresponds to his/her most visited location, and that his/her workplace corresponds to the second most visited location, both identified in terms of placed calls. We performed a sensitivity analysis on this minimal assumption, by imposing in addition some constraints on the time of the call, to refine our identification of locations of residence and workplace (see Text S1 for additional details) [35,38]. We thus define a commuting network at the level of cell sites, creating a directed link between each residence and workplace and assigning a weight equal to the total number of users that commute between the two locations. We coarse-grain the mobile phones commuting network from the tower cell scale to the country's administrative subdivisions for comparison with the census data (see Section 1.3 in Text S1 for additional details).

Once defined on the same geography, the two datasets also need to refer to the same population. The census dataset represents the benchmark, as it comprises the entire population of a country (commuters and non-commuters at a given scale) and its mobility features, whereas the commuting data obtained from the mobile phone dataset is affected by the sampling bias corresponding to the operator's coverage and to the selection of the subset available for the analysis (it only therefore represents a fraction of the total population) and by the algorithm used to identify commuting-like movements. We explored the geographic coverage of the mobile phone dataset for the three countries (see the Analyses subsection for the corresponding methodology adopted). With no additional information on the subset of individuals included in the mobile phone datasets, we opt for a *basic normalization* approach that simply rescales the populations of the mobile phone networks at the administrative unit level by the population sampling ratio $n_i^{mp}/N_i$, where $n_i^{mp}$ is the resident population of region $i$ tracked by the mobile phone dataset and $N_i$ is the resident population of region $i$ according to the official census.

More sophisticated choices can be made to account for the sampling biases in a more accurate way, as discussed in the Discussion section, however they would require additional information that may not be easily available for a large set of countries. Our baseline choice for the basic normalization is motivated by imposing minimal requests on additional metadata that may be needed to correctly calibrate the dataset. With the chosen normalization, the total population assigned to each node of the network (including commuters and non-commuters) is equal in the two systems, whereas the relative fraction of commuters may be different in the two cases.

As a sensitivity analysis, and for further comparison with the radiation model (see following subsection), we also consider a *refined normalization* that assumes the same knowledge required by the radiation model – namely, the total number of commuters per administrative unit. Once normalized to the census population of each given region, this amounts to assume that the mobile phone commuting network has the same number of commuters per region as in the census dataset, the same total population per region, and therefore also the same ratio of commuters vs. non-commuters. Differences may arise in the number and identity of commuting destinations per region of residence, and in the distribution of commuter flows on such directions.

In the following we indicate with $w_{ij}^{mp}$ the normalized flux of commuters from the administrative unit $i$ to the administrative unit $j$ obtained from cell phone activity data using the basic normalization, and with $w_{ij}^{mp*}$ the one obtained with the refined normalization.

**Commuting networks simulated with the radiation model**

We create synthetic commuting networks using the *radiation model* [18]. The model has been specifically developed to reproduce commuter movements and has the additional desirable feature of being parameter-free, i.e. it does not require regression analysis or fit on existing data. These characteristics make it the ideal candidate to generate a synthetic commuting network in absence of empirical data to be fitted. The model is based on a stochastic decision process assigning work locations to each potential commuter, thus determining the daily commuting fluxes across the country. In detail, networks are generated by creating a fully connected topology between country's locations, where the weight of the edge connecting a node $i$ with a node $j$ is defined by the formula [18]:

$$w_{ij}^r = \frac{N_i N_j}{(N_i + P_{ij})(N_i + N_j + P_{ij})} \sum_{j \neq i} w_{ij}, \quad (1)$$

with $N_i$ and $N_j$ being the populations of origin and destination, $P_{ij}$ the total population living between location $i$ and location $j$ (computed as the total population living in a circle of radius $r_{ij}$ centered at $i$, excluding the populations of origin and destination locations), and $\sum_j w_{ij}$ the total number of commuters daily leaving their home in location $i$. Equation (1) assumes the knowledge of population data ($N_i, N_j, P_{ij}$), similarly to what we consider available for the census commuting networks and for the basic normalization of mobile phone mobility commuting fluxes, but it also requires additional information, i.e. the total number of residents who commute in each administrative unit. While the latter information may be easily accessible in developed countries, it is important to note that it may not be routinely collected or available in other regions. Given these quantities, the radiation model yields a commuting flux for each pair $ij$ of administrative units of the country under study; after removing connections having $w_{ij}^r < 1$, a synthetic commuting network at the given resolution scale is obtained. Previous works have already shown the ability of the radiation model to well match census data from the structural and traffic point of view, in a number of countries [18]. Additional comparisons between the radiation model and gravity models have been performed in the UK [39]. Here therefore we do not consider the radiation model in the comparison analysis of commuting networks and discuss instead its adequacy in the framework of spatial epidemic spreading.

## Epidemic metapopulation model

We use a metapopulation modeling approach [40,41] to perform numerical simulations of epidemic scenarios. We assume the national population of every country to be spatially structured in subpopulations defined by the administrative subdivisions described in the previous subsection. We focus on rapid directly transmitted infections, such as influenza-like-illnesses, for which daily regular movements of individuals for commuting purposes were found to well correlate with the observed regional spread [12,13]. We consider a simple SIR compartmental model [41], where individuals can be either susceptible (S), infectious (I) or recovered (R) from the infection, assuming a life-long immunity for recovered individuals. The dynamics is discrete and stochastic and individuals are assumed to be homogeneously mixed within each subpopulation. No additional substructure of the population is considered (e.g. schools or workplaces), as our aim is to introduce a rather simple epidemic model to test the adequacy of different commuting sources for the simulation of ILI dissemination within a country. We therefore neglect unnecessary details that may hinder the interpretation of results. Subpopulations are coupled by directed weighted links representing the commuting fluxes between two locations, thus defining the metapopulation structure of the model [40,41]. No other type of movement is considered.

Human mobility is described in terms of recurrent daily movements between place of residence and workplace so that the infection dynamics can be separated into two components, each of them occurring at each location [42]. The number of newly infected individuals during the working time in location $i$ is randomly extracted from a binomial distribution considering $S_{ii} + \sum_j S_{ji}$ trials (susceptible individuals living and working in location $i$, $S_{ii}$, and susceptible individuals living in $j$ and working in $i$, $S_{ji}$) and a probability equal to the force of infection $\lambda_i^{work} = \beta \frac{(I_{ii} + \sum_j I_{ji})}{(N_{ii} + \sum_j N_{ji})}$ being $\beta$ the transmissibility of the disease, $N_{hk}$ and $I_{hk}$ the total population and the total number of infectious individuals living in location $h$ and working in $k$, respectively. Similarly, the infection events taking place at the resident location during the remaining part of the day are randomly extracted from a binomial distribution considering $S_{ii} + \sum_j S_{ij}$ susceptible individuals and probability equal to the force of infection $\lambda_i^{home} = \beta \frac{(I_{ii} + \sum_j I_{ij})}{(N_{ii} + \sum_j N_{ij})}$. We model an influenza-like-illness transmission characterized by an exponentially distributed infectious period with average $\mu^{-1} = 3$ days [43,44], and explore three epidemic scenarios by varying the transmissibility $\beta$ and corresponding to the following values of the basic reproductive number (average number of secondary cases per primary case in a fully susceptible population [41]): $R_0 = 1.1$, $R_0 = 1.5$, $R_0 = 3.0$, representing a mild, moderate, and severe epidemic, respectively.

Simulations are fully stochastic, individuals are considered as integer units and each process is modeled through binomial and multinomial extractions (more details on the simulation algorithm are reported in Section 4 in Text S1). Each day of the simulation is modeled with commuting movements informed by the three sources considered for a typical working day; therefore no weekends or holidays are envisioned in the model. Simulations are initialized with 10 individuals localized in a given seed. As seeds we consider the country's capital (Lisbon, Madrid and Paris), a peripheral location with a small population (Barrancos, Lleida and Barcelonnette), and a medium size location, characterized by an average population and an average number of connections through commuting links (Braga, Jaen and Rennes). Although the countries under study are geographically contiguous, they are considered as independent entities since the investigated datasets do not include refined data about cross-border commuters. A sensitivity analysis on the role of cross-border commuting in the spread of ILI is reported in Section 3 in Text S1.

Once a set of initial conditions is defined (mobility network, $R_0$, and seeding location), we simulate 1,000 stochastic realizations for each epidemic scenario, for a total duration of 8 months. Such timeframe is chosen as a reference estimate of the expected time comprising the interval from the initial seeding of a pandemic event to the international alert (approximately two months in the case of the 2009 H1N1 pandemic [45]) and the average time period needed to develop a vaccine against the circulating virus (approximately six months) [46]. During this timeframe the value of the basic reproductive number is kept constant, and no change in behavior that could be self-initiated in response to the epidemic [47,48], or imposed by public health interventions is considered, for the sake of clarity in the comparison of the results.

## Analyses

**Coverage of mobile phone dataset**

For each country under study, we assess the coverage of the population in the mobile phone dataset by calculating the national average, $\sum_i n_i^{mp}/N$ (with $N = \sum_i N_i$ being the country population), and the geographic-dependent values at the scale of the administrative units under consideration. By rescaling for the national coverage, we thus measure the ratio $\frac{n_i^{mp}}{N_i} \cdot \frac{N}{\sum_j n_j^{mp}}$ for each region of the countries under study. Values close to 1 would correspond to a geographic distribution of the sample in agreement with the national coverage. The Pearson correlation coefficient is also measured to quantify the correlation between the census population $N_i$ and the rescaled population of mobile phone users $n_i^{mp} \cdot \frac{N}{\sum_j n_j^{mp}}$ across all administrative units.

**Comparison between mobile phone commuting networks and census commuting networks**

We compare the structural and fluxes properties of the commuting networks extracted from census surveys with those of the networks extracted from mobile phones records, in order to test the quality of mobile phones data as a proxy for commuting at national level. We analyze the topology of the networks obtained from the two sources of data and extract the intersection and its associated travel fluxes. We perform different statistical tests (Spearman's rank correlation coefficient, Lin coefficient, and Wilcoxon test) on the correlation between commuting flows connecting any pair of nodes in each dataset, and between the total numbers of commuters per node in each dataset. We also check for non-trivial correlations between the discrepancies found in the two datasets and nodes' populations and distances between connected vertices. The same analysis is run for all countries, at all resolution scales.

**Comparison of the metapopulation epidemic outcomes obtained integrating different mobility sources**

In all realizations and for each subpopulation, we keep track of the following epidemic observables. The temporal information about the epidemic spreading is encoded in the arrival time ($t_a$) of the infection at each subpopulation. The arrival time is defined as the first day an infected individual is recorded (either as a worker or as a resident) in a location with no previously notified cases. The probability distribution of the arrival time and its average value are evaluated for every location. In addition, we discount systematic anticipation/delay effects by subtracting the average arrival time difference $\langle \Delta t_a \rangle$ obtained from the arrival times of all nodes when two different mobility datasets are used (e.g. mobile phone commuting network vs. census commuting network).

The spatial diffusion of the disease is investigated through the epidemic invasion tree representing the most probable transmission route of the infection from one subpopulation to another during the history of the epidemic [14]. In detail, considering a disease-free location $i$, as soon as $I_{ji}(t) \neq 0$ or $I_{ij}(t) \neq 0$ a directed link between $i$ and $j$ is added to the invasion path, meaning that an infectious individual traveled between the two locations importing the infection, or that a susceptible individual acquired the infection at the destination and then returned back to the previously uninfected place of residence. The invasion paths collected from every realization are successively cumulated by assigning to each link a weight equal to the fraction of runs where a certain seeding event has been observed; a minimum spanning tree is finally extracted to obtain the invasion tree.

Since the stochasticity of the seeding events can induce small weights variations in the invasion paths and thus different invasion tree topologies, for every scenario we build 50 invasion trees, each of them obtained from randomly selecting 400 stochastic realizations out of the total of 1,000 run for each scenario (this approach allows us to minimize the random fluctuations in the final invasion tree with a limited computational effort). We then compare the invasion trees describing the spatial spreading on different mobility networks through the Jaccard similarity index. Given a tree $\Gamma_a(\nu_a, \xi_a)$ obtained for scenario $a$ (integrating either the mobile phone commuting network or the radiation model commuting network) identified by $\nu_a$ nodes and $\xi_a$ edges, we calculate the Jaccard index with the tree $\Gamma_c(\nu_c, \xi_c)$ obtained from the census commuting network as $J(\Gamma_a, \Gamma_c) = \frac{\xi_a \cap \xi_c}{\xi_a \cup \xi_c}$, measuring the number of common transmission paths over the total paths. The $J$-value is evaluated between all pairs of invasion trees extracted from the scenarios under comparison on the ensembles of 50 trees per scenario. Average values and reference ranges are calculated.

Incidence and prevalence curves are defined as the density of newly secondary cases and density of infected individuals at every time step. From the ensemble of 1,000 stochastic realizations, average and reference ranges are then evaluated for every location as well as the peak time of the epidemic.

## Results

**Datasets descriptive analysis**

The census commuting networks for Portugal include (i) 1,643,938 commuters traveling between the 278 municipalities through 25,634 weighted directed connections, and (ii) 469,089 commuters traveling between the 18 districts on a fully connected network. In Spain we consider the provinces' geographical scale only, as constrained by the information available in the census survey. The commuting network is formed by 47 nodes and 722 weighted directed edges, representing the daily travel flows of 537,331 commuters. The commuting networks for France are defined at the district scale (8,019,636 commuters moving along 38,077 weighted directed edges connecting 329 nodes), and at the department level (4,957,193 commuters for 7,994 weighted directed links among 96 nodes). For all countries, at all scales considered, all administrative units are included in the datasets (i.e. they have at least one incoming or outgoing commuting flux to another administrative unit in the country). A summary of the basic statistics of the networks extracted from census data is reported in Table 1.

Commuting patterns from mobile phone records are extracted from a sample of 1,058,197 anonymous users in Portugal, 1,034,430 in Spain, and 5,695,974 in France. Records referred to 2,068 towers in Portugal, 9,788 towers in Spain, and 18,461 in France. Once mapped onto the administrative units, we find 452,113, 460,211 and 1,676,103 total commuters in the mobile data samples in Portugal, Spain, and France, respectively, corresponding to the lowest administrative hierarchy.

Population tracked by the operators' samples is distributed nationwide and approximately equal to 9% of the census population in Portugal and France, and 2% of the census population in Spain. By taking into account these scaling factors, cell phone population well correlate with the census population at the highest geographical resolution considered, with a Pearson correlation coefficient between the two quantities equal to $R > 0.9$ ($p < 0.001$) for Spanish provinces, Portuguese municipalities and French districts. Population coverage is rather uniform in France with more than half of the districts in the interval $[0.8 - 1.2]$ of the national coverage value (grey colored units in Figure 1), while larger discrepancies are observed in the geographic distribution of the tracked population in Spain and Portugal. In Spain we observe a significant undersampling of the population in Galicia and Basque regions. In Portugal, we observe larger regional fluctuations around the national coverage value: most of the municipalities report an undersampled population, whereas the region close to the capital, Lisbon, shows an oversampling as large as 3 times the national coverage.

**Statistical comparison of commuting networks**

Commuting networks obtained from census data and mobile phone activity data share the same number of nodes at all hierarchies considered in all countries, given that all administrative units were covered by both datasets, however variations are observed in the number of commuting links (Table 1). The set of links common in both datasets in the Portugal case at the municipality level account for about 60% of the total links of each network and include more than 96% of the total travel flux of both networks. Aggregating the datasets at the level of Portuguese districts, both networks become very close to fully connected, almost achieving a perfect overlap (more than 99% of links falling in the intersection). Similar figures are obtained for French districts, though the common 95% of traffic is distributed over 82% of the census links and only 52% of the mobile phone links. Spain displays a different situation, with the census commuting network topology being completely included into the mobile phone one. Census commuting links represent only 37% of connections of the mobile phone dataset, however accounting for 87% of its total traffic.

We compare the probability density distributions of the travel fluxes $w_{ij}$ in both networks (Figure 2), after considering the basic normalization scaling to the population $N_i$ of each administrative unit (see Methods). All distributions display a broad tail and very similar shapes in each country, and differences are observed in particular for small traffic values. In Portugal and France, the very weak commuting flows are not captured by the mobile phone dataset, clearly as an outcome of the smaller users' sample size in the mobile phones case with respect to census. Such discrepancy disappears when we move to larger spatial scales, as in the case of Spain.

Restricting our analysis on the topological intersection, a side-by-side weight comparison on each link shows a high correlation between the two datasets (Spearman's rank correlation coefficient $> 0.7$ for the largest administrative units, Table 2), however commuting fluxes in the mobile phone network are found to be larger than the census ones across almost the entire interval of values (panels d-f of Figure 2). Deviations appear larger for smaller fluxes ($w_{ij}^c \lesssim 100$ commuters) in Portugal and France, with a good agreement for the largest values, whereas they are uniform in the case of Spain. Similar results are obtained when we analyze the total number of commuters leaving a given administrative unit $i$, as well as the total number of incoming commuters in a given unit. A strong correlation between the two datasets is found for both quantities, generally independent of the level of aggregation considered (Spearman's coefficient $> 0.88$ for Portugal and France), whereas small values of the Lin's coefficient indicate the presence of strong differences in the absolute values for the two datasets ($< 0.53$ across all countries and for all administrative levels, for both quantities, Table 2). Spain has a rather low Spearman's coefficient for the incoming fluxes of commuters with respect to the other countries ($0.54$ vs. values $> 0.88$), showing a poor capacity of the mobile phone data to properly account for the attraction of commuters of a given location.

The correlations found along the various indicators do not ensure the statistical equivalence of the two datasets (a Wilcoxon-test for matched pairs would reject the null hypothesis of zero median differences between paired values of the same quantities).

We further analyze whether the observed discrepancies between the weights in the mobile phone networks and the census networks show any dependency on the variables that characterize the underlying spatial and social structure, namely the Euclidean distance between the connected nodes (calculated from the coordinates of the administrative unit's centroid), the population of the origin node and the population of the destination node (Figure 3). The overestimation of the magnitude of commuting fluxes in the mobile phone dataset does not show a significant dependence on the population sizes. Fluxes are instead found to be more similar when they connect units at shorter distances with respect to longer distances across the countries. Such variation disappears if we consider the topological distance defined by a neighbor joining approach (see Section 3.4 in Text S1). Spatial aggregation into larger administrative units does not alter this overall picture but weakens the effect observed on distance (see details in Section 2.1 in Text S1).

If we refine the normalization of the mobile phone networks by taking into account the total number of commuters in each administrative unit, the agreement with the census dataset improves in the side-by-side weight comparison on every link (see Section 3 in Text S1). This approach allows us to explicitly discount the systematic overestimation found with the basic normalization, resulting in higher

Lin concordance coefficients (Table S1); discrepancies between mobile phone and census data are however still observed for very small and very large commuting flows.

**Epidemic simulations**

We examine whether the observed non-negligible discrepancies in the commuting fluxes of the two datasets are also significant from an epidemic modeling perspective, altering substantially the outcome of disease spreading scenarios. We compare scenarios obtained from stochastic metapopulation models equally defined and initialized, except for the mobility data they integrate (see Methods). In addition to the census commuting network and the mobile phone commuting network, we also consider the synthetic commuting network generated with the radiation model.
Epidemics starting from different seeds in the three countries, and characterized by different values of the basic reproductive number, yield large variations of the Jaccard index value $J$ measuring the similarity in the epidemic invasion paths produced by the use of mobile phone data and of the radiation model with respect to the census benchmark ($J$ in $[0.1,1]$, see Figure 4). Epidemic invasion trees obtained from proxies for mobility are more similar to the ones obtained from the model integrating census data when the seed is located in the capital city of the country. In addition, $J$ increases with larger values of $R_0$.
If the seed is instead located in a peripheral node, values of the Jaccard similarity index fall always below 0.4 in the three countries, and decrease with larger values of the transmissibility.
Mobile phone data performs similarly to the radiation model once the corresponding epidemic models are seeded in a central location, except for the case of Lisbon, and performs better or similar when they are seeded in a peripheral location. If the epidemic starts from a mid-size populated region, the relative performance of the radiation model against mobile phone data in the epidemic outcomes depends on $R_0$, with improvements observed as $R_0$ increases.
To test for the role of overestimation of flows, we also performed the same analysis by considering the refined normalization of the mobile phone commuting data that keeps the same total number of commuters per administrative region as in the census dataset and explicitly discounts overestimation biases. The refined normalization allows the mobile phone data to better reproduce the invasion paths obtained from census commuting flows for central and medium-type locations for all $R_0$, and to perform slightly worse in case the seed is located in a peripheral location (Figure 5 for the case of France).
When focusing on the time of arrival in a given location, we find a systematic difference between models based on proxy networks and the benchmark model integrating census data. Mobile phone data, overestimating the census commuting fluxes if a basic normalization is considered, leads to a positive difference $\Delta t_a = t_a^C - t_a^{MP}$ corresponding to a faster spreading (Figure 4). On the other hand, epidemics on the radiation model tend to unfold slower than simulations on the census network, with later arrival times as indicated by negative values of $\Delta t_a$ (except in the case of France where the median of $\Delta t_a$ is approximately equal to zero in all cases). For small values of $R_0$, the arrival times of simulations running on a proxy network may be substantially different from the ones obtained with census data, with $\Delta t_a$ of the order of months. While the transmission potential of the disease drives the magnitude of the impact of the discrepancies, the role of the seed location appears to be less relevant here than what previously observed in the study of the invasion paths. A slightly decreasing trend in the positive median values of $\Delta t_a$ is observed in the mobile phone vs. census results, going from peripheral to medium to central location, the effect being more pronounced in Spain and in France. By discounting *a posteriori* the average anticipation of the model built on mobile phone data, which is trivially due to the overestimation of the census commuting fluxes, we find a very good correlation between arrival times for the models built on the census network and on the mobile phone network, with most of the points lying close to the identity line (Lin concordance correlation coefficient ranging from 0.77 to 0.88, panels c, f and i of Figure 4). If we consider the refined normalization, anticipation effects produced with the mobile phone data are preserved but reduced in magnitude (Figure 5). Epidemic peak times are also affected by the different distributions of commuting flows in the two networks (see Section 2.2 in Text S1). As soon as the disease reaches most of the nodes, the epidemic model integrating the mobile phone network displays a more homogeneous behavior, with epidemic peaks that follow very shortly after each other in all the subpopulations, while peak times in the census networks span a wider time frame.

On coarser spatial scales (Portuguese districts, French departments), we obtain a higher similarity between simulated results with proxies vs. census (see Section 2.1 in Text S1), closer to the results observed for Spanish provinces. The performance of the epidemic model built on the radiation is noticeably poorer than the mobile phone network if we consider the coarse-grained scale, for all seeds but the capital. The differences between arrival times are generally reduced by the coarse-graining, but remain significant when the reproduction number is small ($\Delta t_a$ ranging between 0 and 120 days).

## Discussion

Next to traditional census sources or transportation statistics, several novel approaches to quantifying human movements have become recently available that increase our understanding of mobility patterns [21-28,49-52]. Adequately capturing human movements is particularly important for improving our ability to simulate the spatiotemporal spread of an emerging disease and enabling advancements in our predictive capacity [53,54]. Previous work has focused on testing mobility models' performance in reproducing the movements of individuals [18,19], and its impact on epidemic simulation modeling results when fully supported by data [19]. The full knowledge of mobility data from national statistics is however largely limited to few regions of the world [14], whereas in many others it may not be routinely collected nor accessible. If mobility models often require aggregated input data from national statistics on movement habits [18] or the full mobility census database [19] for the fitting procedure, mobile phone data may be thought as an ideal alternative candidate for a proxy of human movements in absence of (complete and/or high-resolution) mobility data from official sources [24,26,27].

In order to systematically test this hypothesis exploiting the full resolution of both the proxy data and the official census data for commuting, we have compared these two datasets in three European countries and performed a rigorous assessment of the adequacy of proxy commuting patterns – extracted from mobile phone data or synthetically modeled – to reproduce the spatiotemporal spread of an emerging ILI infection.

Mobility data from mobile phones is able to well capture the fluxes of the commuting patterns of the countries under study, reproducing the large fluctuations in the travel flows observed in the census networks. In all countries the intersection between the two networks includes the vast majority of the commuting flows and the correlation measured on links' traffic and nodes' total fluxes of incoming or outgoing commuters is high (though not statistically equivalent). This suggests that mobile phone data can be used as a surrogate tracking the commuting patterns of a given country, identifying the relative importance of its mobility connections in terms of flows' magnitude, with a resolution that is equivalent to the one adopted by official census surveys or higher. This is a particularly relevant result for data-poor situations, where census data may not be available and official statistics may not be enough to correctly inform a mobility model.

Discrepancies are however found, especially in the overestimation of commuting flows per link and in the larger variations observed for weaker flows and longer distances, that appear to be responsible for the differences observed in the simulated epidemics.

Epidemics run on mobile phone commuting networks well reproduce the simulated invasion pattern on the census commuting when the seed is located in a central location and $R_0$ is large. The capital city is indeed strongly connected to the rest of the country; therefore it behaves as a potential seeder of the direct transmission to the majority of the other cities, leading to very similar star-shaped infection trees from the seed. These rather similar sets of infected locations at the first generation of the invasion path provide a twofold contribution to the increase of $J$: on one side, they correspond to a large fraction of the total number of infected subpopulations, so they contribute a large relative weight in the computation of $J$; on the other, common infected locations are likely to maintain the similarity of the invasion paths at the second generation too, repeating the process in an avalanche fashion. Such behavior becomes increasingly stronger as $R_0$ grows larger.

The opposite situation is instead found when seeds are located in peripheral nodes, reporting low values of the Jaccard index. The analysis of the commuting networks has indeed shown that larger discrepancies exist for small weights. Once considered in the framework of an epidemic propagation, such discrepancies are expected to lead to strong differences in the invasion already at the first generation of infected locations. If these locations directly infected by the seed strongly differ, their contribution to the decrease of the similarity of the invasion paths will become increasingly stronger for further generations: different nodes are infected and likely different neighbors of those nodes will be affected by the disease, so that deviations cumulate at each successive step of the invasion (Figure 6).

Diseases with a higher transmission potential would enhance this behavior, as with a large value of $R_0$ the peripheral seed can more quickly infect a large fraction of the system in the mobile phone network, than in the census dataset. Such effect is also present in the radiation model that is not able to describe the epidemic behavior better than the mobile phone data when the seeding location is characterized by a small population or degree. Not being able to well capture the mobility coupling between peripheral regions and the rest of the country, the radiation model misses most of the seeding events on long distances even when $R_0$ is large (Figure 6). Using a synthetic proxy is therefore not always preferable to data alternatives, and mobile phones appear to be more reliable in matching the spatial epidemic spread starting from peripheral locations.

A clear bias, which is observed consistently across all countries and for all resolution scales considered, is the faster rate of spread of the simulation based on the mobile phone commuting network with respect to the census one. This is clearly induced by the larger commuting flows obtained following the extraction of commuting patterns from mobile phone data using a basic normalization. The effect is stronger for $R_0 = 1.1$ as it is enhanced by the intrinsic large fluctuations characterizing epidemics close to the threshold. In such scenarios, even relatively small differences between networks' topologies can strongly alter the invasion path of the disease, consistently with the results of previous work on the effect of network sampling on simulated outbreaks [53]. Increasing the value of the reproduction number leads to narrower $\Delta t_a$ ranges, because the larger disease transmissibility accelerates the spreading, synchronizing the epidemic behavior at distant locations and, in general, reducing the system's heterogeneity.

Time of arrival of the infection in a given location is better matched by the epidemic model built on the radiation model, though with large fluctuations for small values of $R_0$. However, it is important to keep in mind that the total number of commuters per administrative unit is an input of the radiation model and no overestimation effects, as the ones resulting from the use of mobile phone data in the basic normalization approach, are possible in the model. If we inform the extraction of commuting patterns from mobile phone data with the same input data of the radiation model, i.e. through the refined normalization, predictions on the time of arrival consistently improve with respect to the basic normalization approach. Fixing the total number of commuters equally in the two datasets is however not enough to obtain an equivalent picture in terms of arrival times, as a considerable anticipation for small values of the transmissibility is still observed. These results need to be taken into account when considering epidemic simulations integrating mobility proxies, as a high accuracy in predicting arrival times can be used for assessing the epidemic situation at the source of the infection, estimating important epidemiological parameters during the early phase of the outbreak in a backtracking fashion [11,45,54].

Nodes ranking according to time to first infection also improve in the epidemic simulations based on the refined normalization with respect to the baseline one. The similarity in the invasion paths equals (or even improves) the levels reached once the radiation model is considered. Similar results are therefore obtained from two different sources however employing the same type and amount of input data (for calibration/normalization). Jaccard index values display anyway the presence of important differences in the way the epidemic propagates on proxies with respect to census, being $J > 0.7$ only when the outbreak is seeded in Paris.

Effects of flows overestimation are visible in the analysis of the epidemic peaks too, but less prominent. The larger number of commuters that travel in the mobile phone networks tends to synchronize the epidemic peak between different subpopulations, leading to shorter overall timespan for all subpopulations to peak in the mobile phone case with respect to census. Differences between the datasets mostly range in a time interval of 2-3 weeks, a time resolution that still allows a meaningful comparison of epidemic results with the average reporting period of standard surveillance systems.

In the case of France and Portugal we have also studied multiple hierarchical levels of the administrative units, by aggregating both datasets. Overall, our analysis indicates that the epidemic behavior on aggregated proxy network better matches the results obtained on census data, with respect to higher resolution level. This is however obtained at the cost of studying the epidemic on a lower geographic resolution, which would then provide less information on the predicted time course of the epidemic and may compromise our ability to use models to extract valuable public health information for epidemic control [54]. On the other hand, the radiation model displays an opposite behavior when aggregating on space. This suggests that at each scale of resolution there exists an optimal proxy for the description of the spatial spread of an infectious disease epidemic, similar to what observed in a comparison of mobility models [55].

The overall picture we presented clearly shows that proxies integrated into epidemic models can provide fairly good estimation of the ranking of subpopulations in terms of time to first infection. A good agreement in the simulated arrival times is intrinsically related to proxies' calibration and normalization aspects, and observed biases can be reduced by using additional information, such as the knowledge of the total number of commuters in each location. On the other hand, the most probable path of infection from one subpopulation to another appears to be affected by more substantial discrepancies between the different sources of data or synthetic flows that cannot be overcome through a simple normalization. In order to further improve predictions on the path of invasion, we would need to comprehensively understand the causes behind the differences observed in the data analysis. These are inevitably related to the methods used to account for the population sample considered in the mobile phone data and to define the commuting mobility per user.

First, in extracting the commuting behavior of each user from mobile phone data we necessarily have to make assumptions on the identification of home and work locations (in absence of metadata on the user). If we identify these two locations as the two most visited ones [22], by definition, we are assuming that place of residence and place of work are two distinct locations, yielding that every mobile phone user is a commuter at the resolution level of the cell phone towers. Once aggregated at a larger scale (i.e. the various administrative units under consideration), we obtain a population made of individuals living and working in the same unit (non-commuters) and of individuals commuting between two different units. While aggregation leads to a certain fraction of non-commuters, the resulting commuting behavior – expressed by the ratio of commuters vs. non-commuters – is anyway more pronounced likely because of the intrinsic assumption made on the original identification of home/work locations from the data. Different choices can be made that can improve the correct identification of home/work locations, leveraging on the availability of additional data. If timing of the call activity is provided, one possible refined definition would be to identify as home location the tower cell with the largest activity during nighttime, and the work location as the one with the largest activity constrained to daytime (with variations of the definition of these intervals) [32,33,38,56]. We tested this approach in the Portuguese dataset and found that the identification of the two locations was not substantially altered by the time-constrained definition chosen, and did not affect our results (see Text S1).

Increasingly sophisticated approaches can also be envisioned, based on clustering methods applied to calling behavior [57,58]. In addition to the need for access to the metadata associated to the activity data, results from time-constrained or clustering methods may anyway be affected by biases induced by users' call plans (influencing the pattern of calls to given timeframes during the 24h or depending on the day of the week, e.g. weekday vs. weekend), job types (altering the expected timing pattern of call activity from work), and more generally the definition of normal business hours that may have a strong cultural component.

Second, our basic normalization may be too simplistic, thus inducing strong overestimation because the population sampled through the mobile phone data is not representative of the general population, being characterized by specific different features affecting the resulting mobility behavior. Biases may be induced by mobile phones ownership, with fluctuations strongly dependent on socio-economic status [31,58], and by market share of the specific operator providing the data, In Spain, for example, the strong undersampling of the population in Galicia and Basque region, characterized by a strong political and cultural identity, may be due to the presence of local operators that account for a larger market share than what observed at national level. In Portugal, we observe larger fluctuations per region around the average national coverage than in other countries. Predictions for the invasion path obtained with mobile phones for epidemics starting in the capital of the country are not in good agreement with those obtained with census flows ($J < 0.5$, Figure 4). In this case, the central role of the capital, responsible for leading to higher similarity as discussed before, is reduced by the presence of larger (and overestimated) flows connecting less central regions in the mobile phone dataset. This leads to the creation of leaves stemming from peripheral nodes and infecting the closest neighbors, thus strongly reducing the role of the seed in infecting the large majority of nodes at the first generation of invasion. This phenomenon is effectively similar to the one encountered when the epidemic is seeded in a peripheral location.

Small-scale studies targeting specific populations (such as e.g. a city or a college town) with additional metadata accompanying the activity records may possibly shed more light in the identification of such biases.

In poorer countries these effects are expected to be of a larger magnitude, given that mobile phone users still represent a privileged minority of the population [31]. Recent work has however showed that mobility estimates in Africa are very robust to biases in phone ownership [59].

The introduction of a refined normalization to account for the non-representative nature of the mobile phone sample fixes the total number of commuters equally in the two datasets and leads to an improvement of the comparison of the commuting fluxes on a link-by-link basis. Discrepancies on traffic flows along links are however still observed that are responsible for differences in the resulting epidemic observables, even though the overall systematic overestimation obtained with the basic normalization has been discounted. Increasingly sophisticated approaches can be developed that use iterative proportional fitting, fixing two marginal values that need to be assumed, i.e. the total numbers of incoming commuters and of outgoing commuters per location (or additional data, such as points of interest in the case of intra-city commuting) [60]. Knowledge of these quantities may however not be largely accessible across different regions of the world.

Third, there may be inconsistencies in the definition of commuting for both datasets, or differences in the year of collection of each dataset. We have no information on users' age in the mobile phone dataset, therefore movements for work or study are both tracked in users' trajectories. Commuting for study reasons is included in the Portuguese and French census data, whereas Spain reports about workflows only. The impact of not considering students' commuting in the Spanish case is however estimated to be rather low. Spanish data is indeed collected at a high administrative level (provinces), where students' commuting flows may be very weak given that they are usually more localized than those of workers. Data from France shows that 95% of students (aged<15) travel on distances less than 10km [4]. In order to estimate the impact of missing students in the Spanish dataset on the province scale, we examined the fraction of commuter movements of students in the French census commuting network aggregated at the level of regions, i.e. similar to the size of Spanish provinces. In France, students represent about 10% of the total commuting flows across regions. If we assume a similar statistics for Spain too, such ratio is not sufficient to explain the discrepancy observed between the normalized mobile phone commuting flows and the census commuting flows in Spain (Table 1). In addition, the lack of a portion of individuals in the dataset (e.g. students) would have no impact when using the refined normalization because, in that case, the total number of commuters is set to be equal in both data sets by definition.

Discrepancies in the year of data collection for the two sources range from two years for Spain (2005 is the year of collection of census data, 2007 the year of collection of phone data) to five years for Portugal (2001, 2006). In the case of France, the two datasets belong to the same year (2007). To assess the possible changes in commuting flows with time, we analyzed French yearly data between 2006 and 2009, given their availability (see Table S1). In three years, the total number of commuters in the country increased by about 3%, and every year, the total number of commuters increased by 1% or less. Assuming that similar trends apply to the other countries, we conclude that the total census commuting flows of Spain and Portugal may have increased by 2% and 5%, respectively, a difference being much smaller than the average discrepancy observed between census data and mobile phone data.

Finally, the epidemic model considered adopts some approximations that we would like to discuss in the following. Even if countries under consideration belong to a contiguous area in continental Europe, numerical simulations for the epidemic spread were performed for each country in isolation. This choice is driven by the lack of mobile phone data for cross-border movements (given their national nature), and by the negligible fraction of commuting across countries with respect to national commuting (about 780,000 people in the EU, including EEA/EFTA, were cross-border commuters in the year 2006/2007 [61] over a total of more than 100 million national commuters). For the sake of completeness, we also checked our results against the inclusion of cross-border commuting in the census network of France, where international movements are predominant in a subset of districts (for instance, in those bordering Switzerland). Results reported in Text S1 show that including cross-border movements in the census commuting networks does not significantly alter the simulated epidemic patterns, keeping our conclusions unchanged.

The modeling approach we proposed was fairly simple and did not consider additional substructure of the population, interventions, change of behavior or weekend vs. weekday movements. Our aim not being to reproduce historical epidemics, we chose to include only the basic ingredients that were the object of the analysis in order to achieve a clearer understanding and interpretation of results. Simulations were performed assuming a continuous series of working days, given the purpose of the study and the knowledge that the inclusion of weekend movements has little or no effect in the resulting epidemic profile [42]. To apply this framework to real case studies, more refined compartmental models, movements and interactions between individuals may need to be considered. Our study was performed on three European countries, and we expect that our conclusions are applicable to other developed countries in the world characterized by similar cultural, social, and economic profiles.

Our approach for the extraction of commuting patterns from mobile phone data was based on minimal assumptions in order to facilitate its generalizability in other settings where data knowledge may be limited or completely absent. Further work is necessary to extend this work to the analysis of the adequacy of mobile phone data as proxy for human mobility in underdeveloped countries where cultural and socio-economic factors may affect differently the biases here exposed. We also note that diseases other than ILI may be of higher interest for these regions, and in that case the relevant mobility mode and epidemic model would need to be updated in the approach we presented.

For instance, the transmission of the disease under study may be strongly affected by seasonal forces, such as the variations in human density and contact rates due to agricultural cycles that drive the spatial spread of measles in Niger [62], or other factors, such as poor sanitation standards that were linked to the persistence of the poliovirus in India [63]. Also, long-term migration may play a more significant role than commuting movements in the spread of the polio virus. On the other hand, the concern for the emergence of new infectious diseases with pandemic potential, as in the recent cases of the H7N9 flu in China [64] and the MERS-CoV virus in the Middle East [65,66], is significant for developing countries as well, given they may have access to fewer resources for preparedness and control. In this context, our work can provide useful insights for the development of epidemic models for the spatial spread of such rapidly disseminated directly transmitted emerging diseases.

## Acknowledgements

GKKK would like to thank the ISI Foundation in Turin for its hospitality during the time this work was undertaken.

## References

1. Eurostat, Tourism Statistics. Available at: http://epp.eurostat.ec.europa.eu/portal/page/portal/tourism/introduction
2. United States Census Bureau, Commuting (Journey to Work). Available at: http://www.census.gov/hhes/commuting/
3. UK Data Service Census, Census Flow Data. Available at: http://census.ukdataservice.ac.uk/get-data/flow-data.aspx
4. Institut national de la statistique et des études économiques, Bases sur les flux de mobilité: mobilités professionnelles. Available at: http://www.insee.fr/fr/bases-de-donnees/
5. Statistics Portugal. Available at: http://www.ine.pt
6. Instituto Nacional de Estadística, Encuesta de población activa. Available at: http://www.ine.es/inebaseDYN/epa30308/epa_inicio.htm
7. Garske T, Yu H, Peng Z, Ye M, Zhou H, et al. (2011) Travel Patterns in China. PLoS ONE 6(2): e16364. doi:10.1371/journal.pone.0016364
8. Colizza V, Barrat A, Barthélemy M, Vespignani A (2007) Predictability and epidemic pathways in global outbreaks of infectious diseases: the SARS case study. BMC Med 5, 34. doi:10.1186/1741-7015-5-34
9. Hufnagel L, Brockmann D, Geisel T (2004) Forecast and control of epidemics in a globalized world. Proc Natl Acad Sci USA 101: 15124 – 15129. doi:10.1073/pnas.0308344101
10. Kahn K, Arino J, Hu W, Raposo P, Sears J et al. (2009) Spread of a Novel Influenza A (H1N1) Virus via Global Airline Transportation. N Engl J Med 361: 212-214. doi:10.1056/NEJMc0904559
11. Balcan D, Hu H, Gonçalves B, Bajardi P, Poletto C, et al. (2009) Seasonal transmission potential and activity peaks of the new influenza A(H1N1): a Monte Carlo likelihood analysis based on human mobility. BMC Med 7: 45. doi:10.1186/1741-7015-7-45
12. Viboud C, et al. (2006) Synchrony, waves, and spatial hierarchies in the spread of influenza. Science, 312**:** 447-451. doi:10.1126/science.1125237
13. Charaudeau S, Pakdaman K, Boëlle P-Y (2014) Commuter Mobility and the Spread of Infectious Diseases: Application to Influenza in France. PLoS ONE 9(1): e83002. doi:10.1371/journal.pone.0083002
14. Balcan D, Colizza V, Gonçalves B, Hu H, Ramasco JJ, et al. (2009) Multiscale mobility networks and the large scale spreading of infectious diseases, Proc Natl Acad Sci USA 106: 21484-21489. doi:10.1073/pnas.0906910106

15. Ciofi degli Atti ML, Merler S, Rizzo C, Ajelli M, Massari M, et al. (2008) Mitigation Measures for Pandemic Influenza in Italy: An Individual Based Model Considering Different Scenarios. PLoS ONE 3(3): e1790. doi:10.1371/journal.pone.0001790.
16. Ferguson NM, Cummings DAT, Cauchemez S, Fraser C, Riley S, et al. (2005) Strategies for containing an emerging influenza pandemic in Southeast Asia. Nature 437: 209 – 214. doi:10.1038/nature04017
17. Ortúzar J deD, Willumsen LG (2001) Modelling Transport, Fourth Edition. Wiley. 606p. doi:10.1002/9781119993308
18. Simini F, González MC, Maritan A, Barabási A-L (2012) A universal model for mobility and migratory patterns. Nature 484: 96 – 100. doi:10.1038/nature10856
19. Truscott J, Ferguson NM (2012) Evaluating the Adequacy of Gravity Models as a Description of Human Mobility for Epidemic Modelling. PLoS Comput Biol 8(10): e1002699. doi:10.1371/journal.pcbi.1002699.
20. Merler S, Ajelli M. (2009) The role of population heterogeneity and human mobility in the spread of pandemic influenza. Proc. R. Soc. B. 277(1681): 557 – 565. doi: 10.1098/rspb.2009.1605
21. González MC, Hidalgo CA, Barabási A-L (2008) Understanding individual human mobility patterns. Nature 453: 779-782.
22. Song C, Qu Z, Blumm N, Barabási A-L (2010) Limits of predictability in human mobility. Science 327: 1018–1021.
23. Song C, Koren T, Wang P, Barabási A-L (2010) Modelling the scaling proper- ties of human mobility. Nat Phys 6: 818–823.
24. Wesolowski A, Eagle N, Tatem AJ, Smith DL, Noor AM et al. (2012) Quantifying the impact of human mobility on malaria. Science 338: 267-270.
25. Bengtsson L, Lu X, Thorson A, Garfield R, von Schreeb J (2011) Improved response to disasters and outbreaks by tracking population movements with mobile phone network data: a post-earthquake geospatial study in Haiti. PLoS Med 8: e1001083
26. Tatem AJ, Qiu Y, Smith DL, Sabot O, Ali AS, Moonen B (2009) The use of mobile phone data for the estimation of the travel patterns and imported Plasmodium falciparum rates among Zanzibar residents. Malaria Journal 8:287.
27. Le Menach A, Tatem AJ, Cohen JM, Hay SI, Randell H et al. (2011) Travel risk, malaria importation and malaria transmission in Zanzibar. Sci. Rep. 1: 93.
28. Lu X, Bengtsson L, Holme P (2012) Predictability of population displacement after the 2010 Haiti earthquake. Proc Natl Acad Sci USA 109: 11576-81.

29. Lima A, De Domenico M, Pejovic V, Musolesi M (2013) Exploiting Cellular Data for Disease Containment and Information Campaigns Strategies in Country-Wide Epidemics. Third International Conference on the Analysis of Mobile Phone Datasets (NETMOB'13). Boston, USA. May 2013.
30. Wesolowski A, Eagle N, Noor AM, Snow RW, Buckee CO (2012) Heterogeneous Mobile Phone Ownership and Usage Patterns in Kenya. PLoS ONE 7(4): e35319. doi:10.1371/journal.pone.0035319
31. Blumenstock, JE, Eagle N (2012) Divided We Call: Disparities in Access and Use of Mobile Phones in Rwanda. Information Technology and International Development, 8(2): 1-16.
32. de Montjoye YA, Hidalgo CA, Verleysen M, Blondel VD (2013) Unique in the Crowd: The privacy bounds of human mobility. Sci Rep 3: 1376. doi: 10.1038/srep01376.
33. Bharti N, Xia Y, Bjornstad ON, Grenfell BT (2008) Measles on the Edge: Coastal Heterogeneities and Infection Dynamics. PLoS ONE 3(4): e1941. doi:10.1371/journal.pone.0001941
34. Calabrese F, Smoreda Z, Blondel VD, Ratti C (2011) Interplay between Telecommunications and Face-to-Face Interactions: A Study Using Mobile Phone Data. PLoS ONE 6(7): e20814. doi:10.1371/journal.pone.0020814
35. Phithakkitnukoon S, Smoreda Z, Olivier P (2012) Socio-Geography of Human Mobility: A Study Using Longitudinal Mobile Phone Data. PLoS ONE 7(6): e39253. doi:10.1371/journal.pone.0039253
36. Phithakkitnukoon S, Leong TW, Smoreda Z, Olivier P (2012) Weather Effects on Mobile Social Interactions: A Case Study of Mobile Phone Users in Lisbon, Portugal. PLoS ONE 7(10): e45745. doi:10.1371/journal.pone.0045745
37. Sobolevsky S, Szell M, Campari R, Couronné T, Smoreda Z, et al. (2013) Delineating Geographical Regions with Networks of Human Interactions in an Extensive Set of Countries. PLoS ONE 8(12): e81707. doi:10.1371/journal.pone.0081707
38. Schneider CM, Belik V, Couronné T, Smoreda Z, González MC (2013) Unravelling daily human mobility motifs. J R Soc Interface 10: 1742. doi: 10.1098/rsif.2013.0246
39. Masucci AP, Serras J, Johansson A, Batty M (2013) Gravity versus radiation models: On the importance of scale and heterogeneity in commuting flows. Phys Rev E 88: 022812.
40. Hanski I, Gilpin ME (1997) Metapopulation Biology: Ecology, Genetics, and Evolution. Academic Press, San Diego.
41. Keeling MJ, Rohani P (2008) Modeling infectious diseases in humans and animals. Princeton University Press.

42. Danon L, House T, Keeling MJ (2009) The role of routine versus random movements on the spread of disease in Great Britain. Epidemics 1: 250–258.
43. Longini IM Jr, Halloran ME, Nizam A, Yang Y (2004) Containing pandemic influenza with antiviral agents. Am J Epidemiol 159: 623-633.
44. Longini IM Jr, Nizam A, S X, Ungchusak K, Hanshaoworakul W, Cummings D, Halloran ME (2005) Containing pandemic influenza at the source. Science 309: 1083-1087.
45. Fraser C, Donnelly CA, Cauchemez S, Hanage WP, Van Kerkhove MD, Hollingsworth TD et al. (2009) Pandemic potential of a strain of influenza A(H1N1): early findings. Science 324: 1557.
46. Abelin A, Colegate T, Gardner S, Hehme N, Palache A (2011) Lessons from pandemic influenza A(H1N1): The research-based vaccine industry's perspective. Vaccine 29, 6: 1135–1138. doi:10.1016/j.vaccine.2010.11.042.
47. Funk S, Salathé M, Jansen VA (2010) Modelling the influence of human behaviour on the spread of infectious diseases: a review. J Roy Soc Interface 7(50):1247-1256. doi:10.1098/rsif.2010.0142
48. Meloni S, Perra N, Arenas A, Gomez S, Moreno Y, Vespignani A (2011) Modeling human mobility responses to the large-scale spreading of infectious diseases. Sci Rep 1:62. doi:10.1038/srep00062
49. Stoddard ST, Morrison AC, Vazquez-Prokopec GM, Paz Soldan V, Kochel TJ, et al. (2009) The role of human movement in the transmission of vector-borne pathogens. PLoS Negl Trop Dis 3(7) : e481. doi:10.1371/journal.pntd.0000481.
50. Vazquez-Prokopec GM, Stoddard ST, Paz-Soldan V, Morrison AC, Elder JP, et al. (2009) Usefulness of commercially available GPS data-loggers for tracking human movement and exposure to dengue virus. Int J Health Geogr 8:68. doi:10.1186/1476-072X-8-68
51. Bharti N, Tatem AJ, Ferrari MJ, Grais RF, Djibo A, et al. (2011) Explaining seasonal fluctuations of measles in Niger using nighttime lights imagery. Science 334: 1424–1427.
52. Lu X, Wetter E, Bharti N, Tatem AJ, Bengtsson L (2013) Approaching the Limit of Predictability in Human Mobility. Sci Rep 3:2923. doi:10.1038/srep02923
53. Dalziel BD, Pourbohloul B, Ellner SP (2013) Human mobility patterns predict divergent epidemic dynamics among cities. Proc R Soc B Biol Sci 280(1766): 1471-2954. doi: 10.1098/rspb.2013.0763.
54. Tizzoni M, Bajardi P, Poletto C, Ramasco JJ, Balcan D, et al. (2012) Real-time numerical forecast of global epidemic spreading: case study of 2009 A/H1N1pdm. BMC Medicine 10:165.

55. Cross PC, Caillaud D, Heisey DM (2012) Underestimating the effects of spatial heterogeneity due to individual movement and spatial scale: infectious disease as an example. Landscape Ecol, 28: 247-257.
56. Isaacman S, Becker R, Caceres R, Kobourov S, Martonosi M, Rowland J, Varshavsky A (2011). Identifying Important Places in People's Lives from Cellular Network Data. In: Lyons K, Hightower J, Huang EM, editors. Pervasive Computing, 9th International Conference, Pervasive 2011, Proceedings. Springer Berlin Heidelberg. pp. 133-151.
57. Csaji BC, Browet A, Traag VA, Delvenne J-C, Huens E, et al. (2013) Exploring the mobility of mobile phone users. Physica A 392:1459.
58. Soto V, Frias-Martinez V, Virseda J, Frias-Martinez E (2011) Prediction of Socioeconomic Levels Using Cell Phone Records. In: Konstan JA, Conejo R, Marzo JL, Oliver N, editors. User Modeling, Adaption and Personalization. 19th International Conference UMAP Proceedings. Springer Berlin Heidelberg. pp. 377-388.
59. Wesolowski A, Eagle N, Noor AM, Snow RW, Buckee CO (2013) The impact of biases in mobile phone ownership on estimates of human mobility. J R Soc Interface 10: 20120986.
60. Bishop YMM, Fienberg SE, Holland PW (1975) Discrete Multivariate Analysis: Theory and Practice. MIT Press.
61. MKW Wirtschaftsforschung GmbH & Empirica Kft (2009) Scientific Report on the Mobility of Cross-Border Workers within the EU-27/EEA/EFTA Countries. Final report to the European Commission. Available at: http://ec.europa.eu/social/BlobServlet?docId=3459&langId=en
62. Ferrari MJ, Djibo A, Grais RF, Bharti N, Grenfell BT, et al. (2010) Rural–urban gradient in seasonal forcing of measles transmission in Niger. Proc R Soc B Biol Sci 277: 2775–2782. doi: 10.1098/rspb.2010.0536
63. Grassly NC, Fraser C, Wenger J, Deshpande JM, Sutter RW, et al. (2006) New Strategies for the Elimination of Polio from India. Science 314: 1150-1153.
64. Gao R, Cao B, Hu Y, Feng Z, Wang D, et al. (2013) Human infection with a novel avian-origin influenza A (H7N9) virus. N Engl J Med 368(20):1888-97. doi:10.1056/NEJMoa1304459
65. Corman V, Eckerle I, Bleicker T, Zaki A, Landt O, Eschbach-Bludau M, et al. (2012) Detection of a novel human coronavirus by real-time reverse-transcription polymerase chain reaction. Euro Surveill 217: 20285
66. Poletto C, Pelat C, Levy-Bruhl D, Yazdanpanah Y, Boelle P-Y, et al. (2014) Assessment of the MERS-CoV epidemic situation in the Middle East region. Euro Surveill (in press).

**Figure Legends**

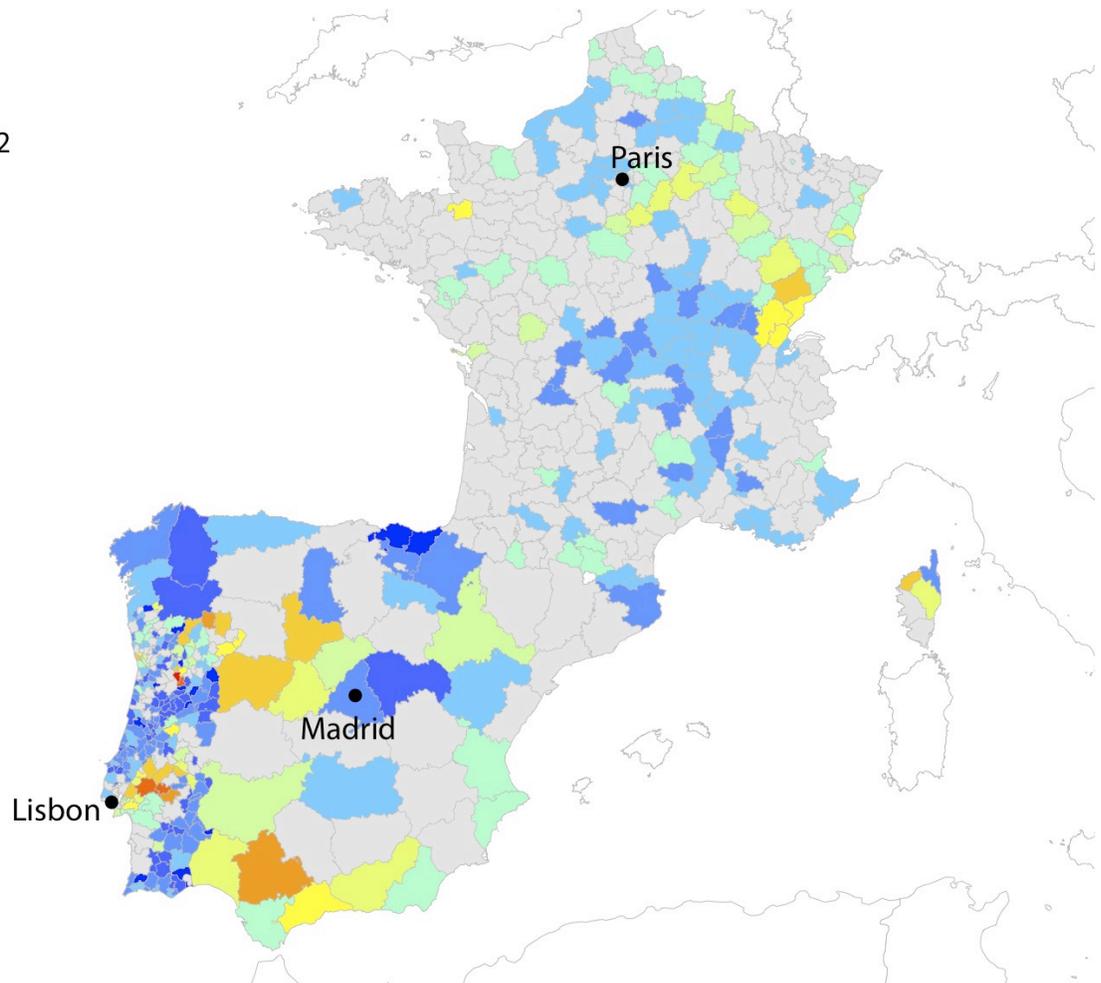

**Figure 1. Spatial differences in coverage of the mobile phones and census datasets.** Map showing the ratio $N_i^{MP}/N_i$ for each region $i$ of the countries under study. $N_i^{mp}$ indicates the population in the mobile phone dataset estimated through equation 3, and $N_i$ represents the official census population. Values close to unity (in grey) indicate that the coverage of the mobile phone dataset is similar to the national coverage; larger (in red) or smaller values (in blue) indicate that the mobile phone dataset is over or under sampling those regions, respectively, compared to the national average. The map was made exclusively for this manuscript and is not subject to copyright.

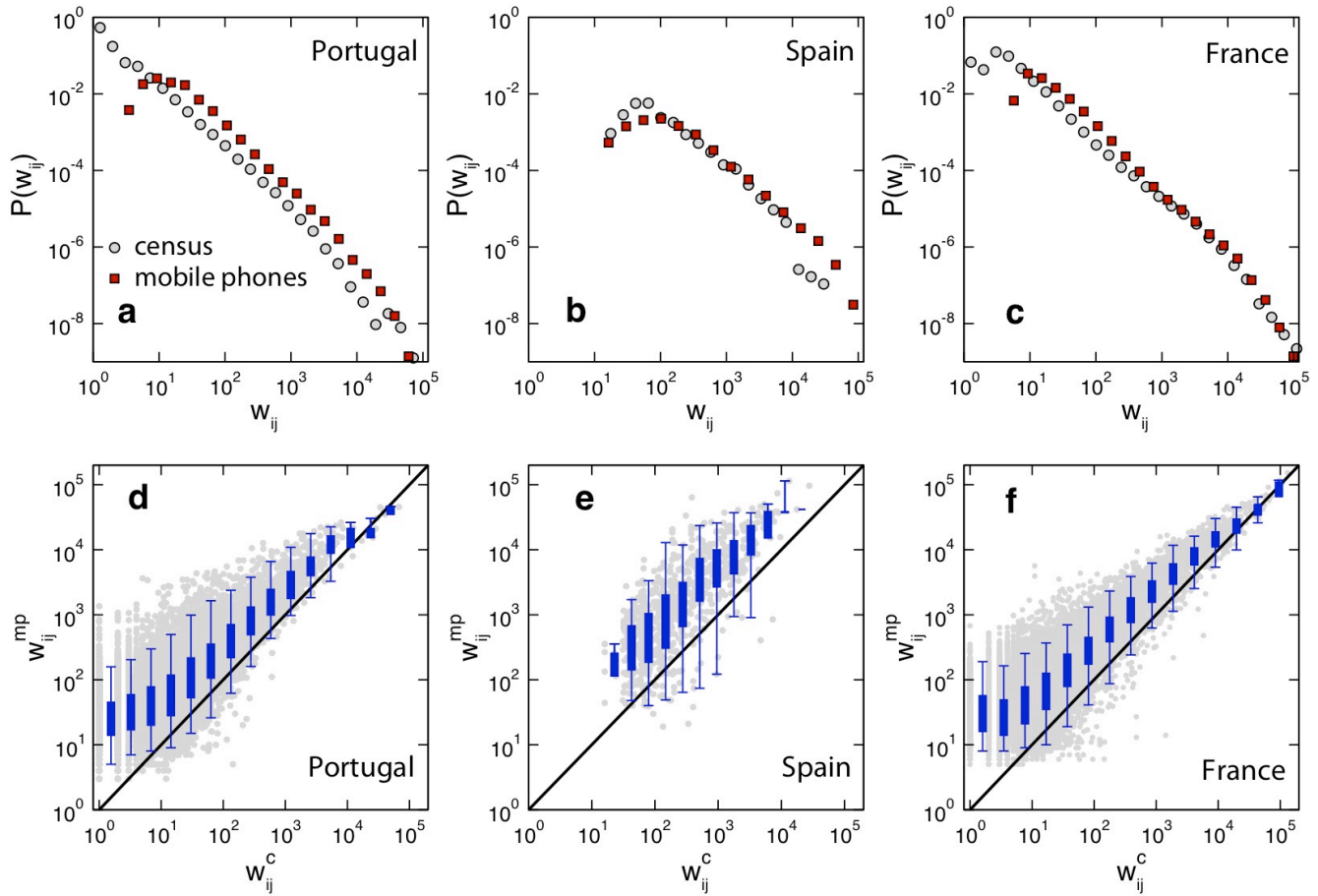

**Figure 2. Comparing the weights of the census networks and the mobile phone networks.** Top: probability density distributions of the weights ($w_{ij}$) of the census commuting network (grey) and the mobile phone commuting network (red) in Portugal (a), Spain (b) and France (c). Bottom: comparing weights in the mobile phone network ($w^{mp}$) and weights in the census networks ($w^c$) in Portugal (d), Spain (e) and France (f). Grey points are scatter plot for each connection. Box plots indicate the 95% reference range of values within a bin.

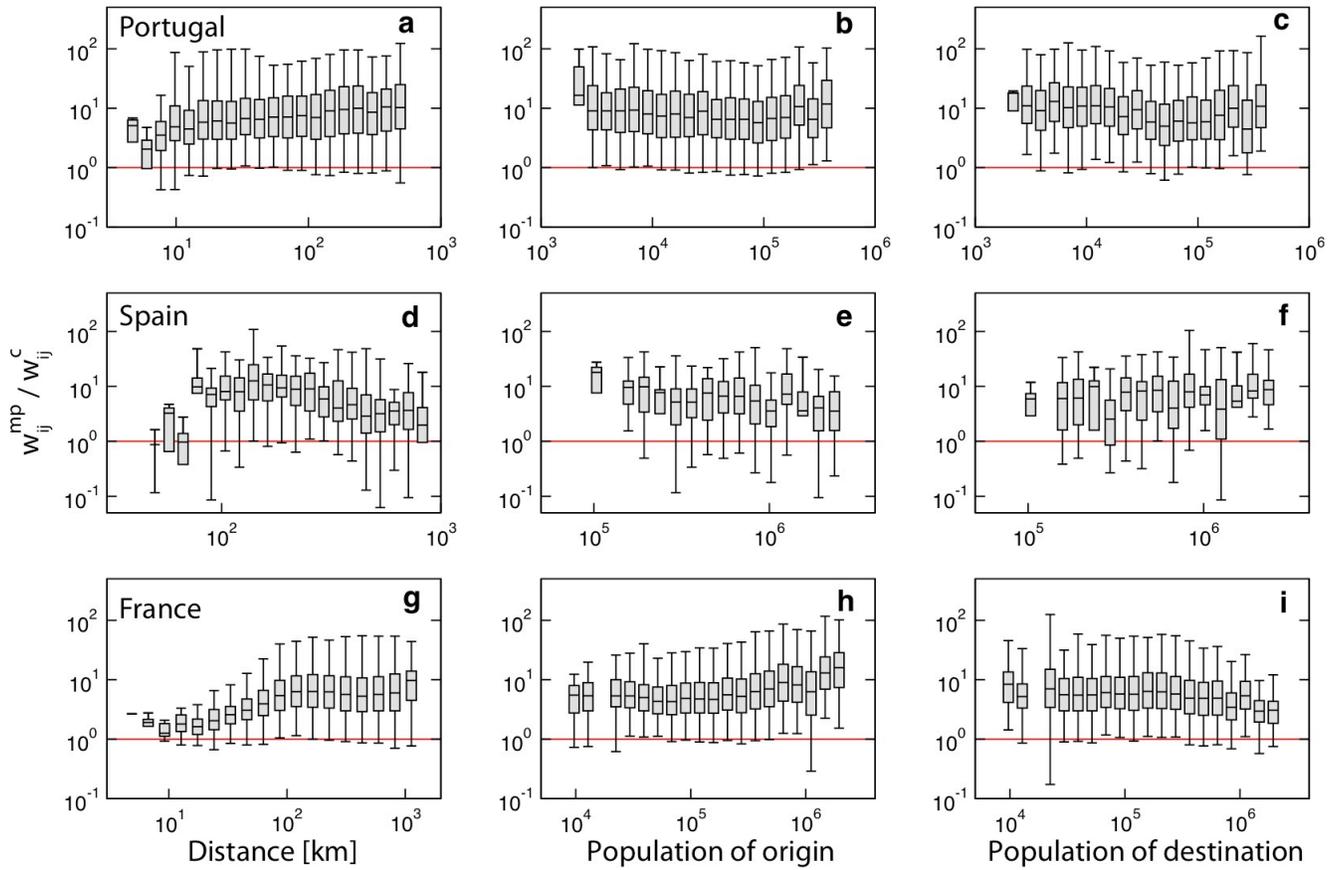

**Figure 3. Effects of geography and demography on commuting fluxes.** Panels show the ratio between the weights of the mobile phone networks $w^{mp}$ and the census networks $w^c$ in Portugal (top panels), Spain (middle) and France (bottom), as function of the Euclidean distance between nodes (a, d and g), the population of origin (b, e, and h) and the population of destination (c, f and i). The solid red line indicates the unit value.

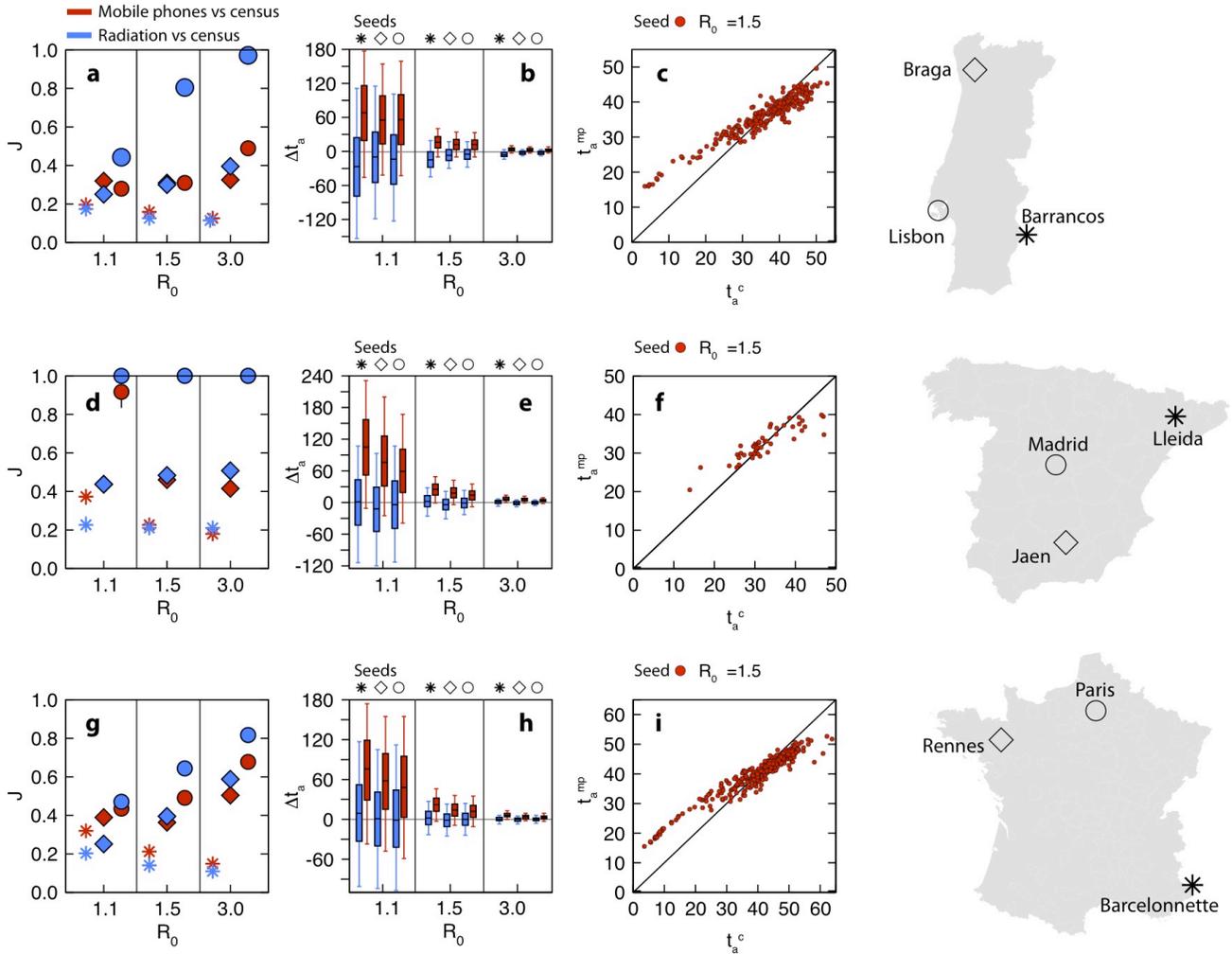

**Figure 4. Epidemic spreading.** Comparing the epidemic behavior on the census network and two proxy networks, mobile phone (red symbols) and radiation model (blue symbols), in Portugal (top panels), Spain (middle) and France (bottom). Panels a, d and g: Jaccard similarity index measured between the epidemic infection tree of the census network and the infection tree of the proxy network, for three values of the basic reproduction number $R_0$. Each symbol corresponds to a different initial infection seed, displayed on the map (right panels). Panels b, e and h: differences between the arrival times in the census network and in the proxy network, for different values of $R_0$ and infection seed. Box plots indicate the 90% reference range, measured on all the network nodes. Panels c, f and i: comparing the arrival times in the mobile phone network $t_a^{mp}$ with those in the census network $t_a^c$, for $R_0=1.5$ and the epidemic starting from the capital city. Red points are scatter plot for each node of the network and we subtracted the average systematic difference $\langle \Delta t_a \rangle$ from each $t_a^{mp}$.

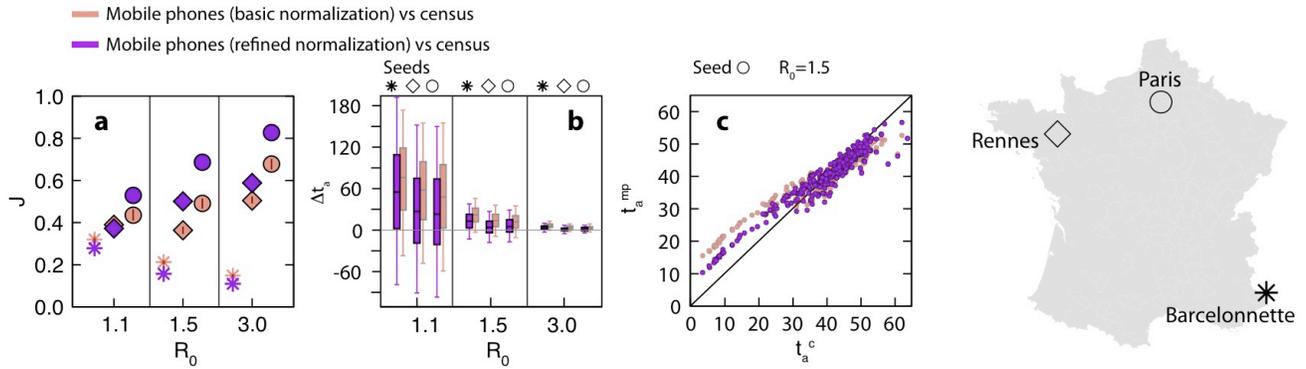

**Figure 5. Epidemic spreading considering the refined normalization.** Comparing the epidemic behavior of mobile phone proxy vs. census, when basic and refined normalization are considered. Only the case of France is shown. (a): Jaccard similarity index of the epidemic infection tree. Each symbol corresponds to a different initial infection seed, displayed on the map (on the right). (b): Differences between the arrival times in the census network and in the proxy network, for different values of $R_0$ and infection seed. Box plots indicate the 90% reference range, measured on all the network nodes. (c): Arrival times in the mobile phone network $t_a^{mp}$ compared with those in the census network $t_a^c$, for $R_0=1.5$ and the epidemic starting from the capital city. Red points are scatter plot for each node of the network and we subtracted the average systematic difference $\langle \Delta t_a \rangle$ from each $t_a^{mp}$.

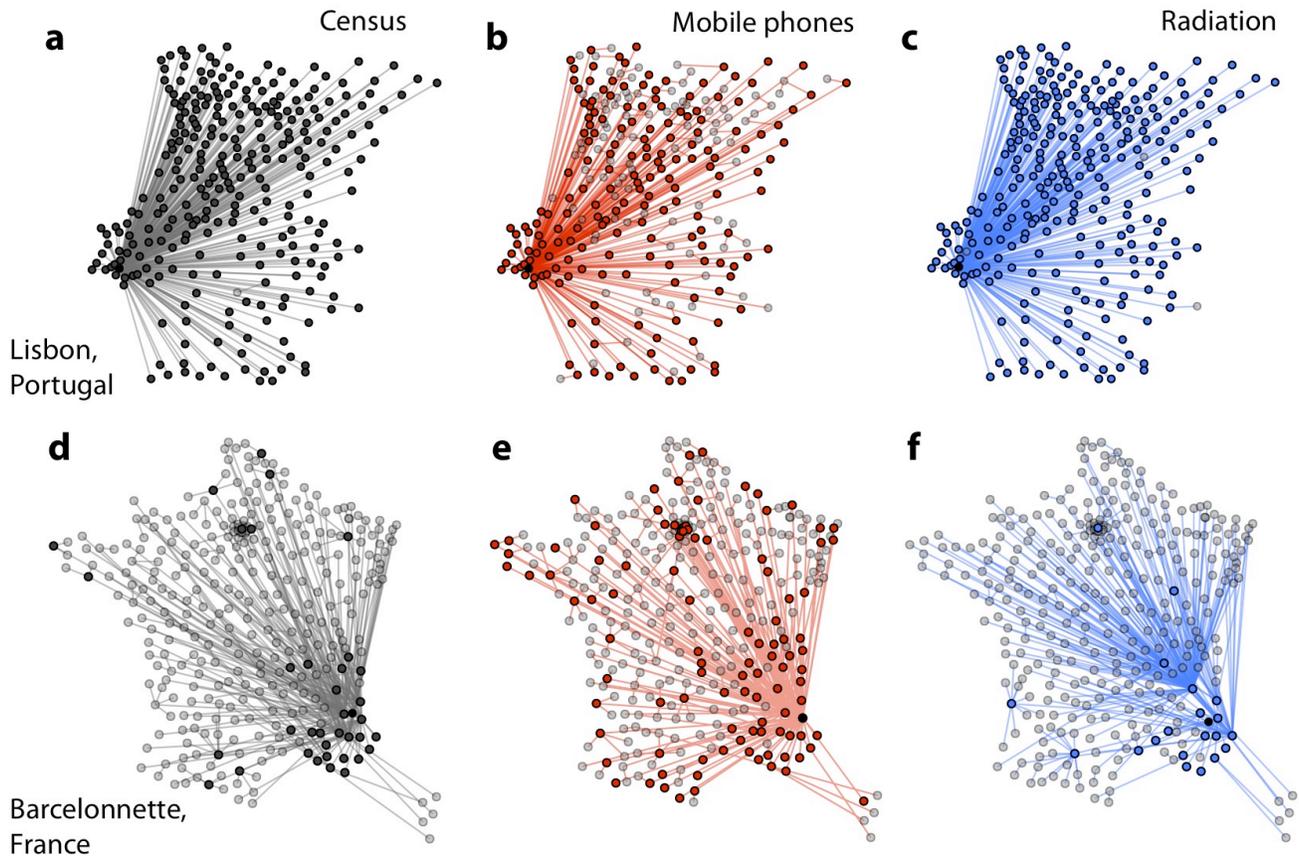

**Figure 6. Epidemic invasion trees.** The full invasion trees for $R_0 = 3$ are shown for Portugal (top row) and France (bottom row) in the cases of the census network (a, d), the mobile phone network (b, e) and the radiation network (c, f). Seeds of the simulations (black nodes) are Lisbon for Portugal and Barcelonnette for France. Nodes belonging to the first shell of the tree, i.e. those directly infected from the seed are fully colored. Grey nodes have been infected by secondary infected nodes.

## Supporting Information Legends

**Text S1.** The file contains: additional information on data sources (Section 1). Additional results for lower geographic resolutions and epidemic peak times (Section 2). Sensitivity analysis on cross-border commuting, refined definitions of workplace and residence, refined normalization of phone data (Section 3). Additional details on the simulation algorithm (Section 4).

# Tables

| country | administrative level | # nodes | | # links | | # commuters | | |
|---|---|---|---|---|---|---|---|---|
| | | *census* | *mobile phones* | *census* | *mobile phones* | *census* | *mobile phones* | *mobile phones normalized* |
| Portugal | municipalities | 278 | 278 | 25,634 | 24,846 | 1,643,398 | 452,113 | 5,255,010 |
| | districts | 18 | 18 | 305 | 306 | 469,089 | 155,137 | 3,525,367 |
| Spain | provinces | 47 | 47 | 722 | 2,146 | 537,331 | 460,211 | 5,181,570 |
| France | districts | 329 | 329 | 38,077 | 60,816 | 8,019,636 | 1,676,103 | 18,750,497 |
| | departments | 96 | 96 | 7,994 | 8,930 | 4,957,193 | 1,087,856 | 12,198,666 |

**Table 1. Basic properties of the commuting networks.** Number of nodes, of links, and of commuters for each commuting network under study, without considering self-loops. Rows correspond to different countries and geographical subdivisions within a country. Columns indicate values from the census dataset and the mobile phone dataset. Commuters for the mobile phone dataset correspond to the values obtained directly from the samples, prior to the normalization procedure, and after the basic normalization procedure. Values obtained with the refined normalization are not reported as they are equal to those of the census dataset, by definition.

| country | administrative level | $w_{ij}$ | | outgoing commuters per residence location | | incoming commuters per destination | |
|---|---|---|---|---|---|---|---|
| | | *Lin* | *Spearman* | *Lin* | *Spearman* | *Lin* | *Spearman* |
| Portugal | municipalities | 0.42 | 0.64 | 0.41 | 0.93 | 0.41 | 0.88 |
| | districts | 0.44 | 0.89 | 0.17 | 0.93 | 0.17 | 0.90 |
| Spain | provinces | 0.45 | 0.73 | 0.15 | 0.72 | 0.15 | 0.54 |
| France | districts | 0.53 | 0.67 | 0.56 | 0.93 | 0.53 | 0.95 |
| | departments | 0.49 | 0.81 | 0.47 | 0.93 | 0.41 | 0.94 |

**Table 2. Statistical comparison between census and mobile phone data.** Values of the Lin's concordance coefficients after a log transformation of variables, and Spearman's coefficient measured between the mobile phone network and the census network for the weights ($w_{ij}$) and the nodes' total fluxes for incoming and outgoing commuters. Rows correspond to different countries and geographical subdivisions within a country.